\documentclass[showpacs,preprintnumbers,amsmath,amssymb]{revtex4}

\usepackage{epsfig}
\usepackage{graphicx}% Include figure files
\usepackage{dcolumn}% Align table columns on decimal point
\usepackage{bm}% bold math

\begin{document}

\title{Granger causality for circular variables}
\author{
Leonardo Angelini, Mario Pellicoro, and Sebastiano Stramaglia}
\affiliation{Istituto Nazionale di Fisica Nucleare, Sezione di Bari,
Italy \\
Dipartimento di Fisica, University of Bari, Italy \\
}

\date{\today}% It is always \today, today,

\begin{abstract}
In this letter we discuss  use of Granger causality to the analyze
systems of coupled circular variables, by modifying a recently
proposed method for multivariate  analysis of causality. We show the
application of the proposed approach on several Kuramoto systems, in
particular one living on networks built by preferential attachment
and a model for the transition from deeply to lightly anaesthetized
states. Granger causalities describe the flow of information among
variables. 
\end{abstract}

\maketitle Countless physical systems are effectively represented
using circular variables, such as phases or orientations
\cite{winfree}. Coupled oscillators systems describe, e.g., arrays
of Josephson junctions, chemical reaction diffusion systems,
circadian oscillations, brain activity, and many others
\cite{bocca}. A problem of particular interest is to assess the
interaction between sub-systems, each described by a phase variable,
a task which has been tackled using the ideas of generalized
synchronization, phase synchronization and phase dynamics modeling
\cite{rosen}. A great deal of attention has been recently paid to
the interplay between the properties of the coupling network and
synchronization of oscillators \cite{oscillatornetworks}.

Granger \cite{hla} proposed a major approach to analyze causality
between two time series: if the prediction error of the first time
series is reduced by exploiting the knowledge of the  second one,
then the second time series is said to have a causal influence on
the first one. Initially developed for econometric applications,
Granger causality has gained popularity also among physicists (see,
e.g., \cite{chen,blinoska,seth,faes}). Being closely related to
transfer entropy \cite{schreiber}, Granger causality is connected to
the amount of information being transferred from one time series to
the other. A novel approach for multivariate Granger causality has
been recently proposed in \cite{noi}: the problem of
false-causalities is addressed by a selection strategy of the
eigenvectors of a reduced Gram matrix whose range represents the
additional features due to the inclusion of new variables.

In this contribution we propose use of Granger causality to analyze
systems of coupled oscillators. To this aim, we adapt the approach
of \cite{noi} to handle circular variables; we show, by means of
several examples, that the interpretation of Granger causality, in
terms of flow of information, is sound also in this case.

Let us consider a system of $n$ interacting circular variables
$\theta_1(t)$, $\theta_2(t)$, $\ldots$, $\theta_n(t)$ discretely
sampled in time. First of all we specify the input and output
variables of the regression model. Inspired by previous works
concentrating on detecting the direction of coupling in interacting
oscillators \cite{rosen}, we proceed as follows.

For $t=1,\ldots,N$, $N$ being the number of samples, we call
$\Delta\theta_i(t)$ the phase increments
$\theta_i(t)-\theta_i(t-1)$, and we consider the problem of
predicting $\Delta\theta_i$ on the basis of
$\{\theta_j(t-\delta)\}_{j=1,\ldots,n;\delta=1,\ldots,m}$, i.e. all
the phase values with lags from one to $m$. A causal relationship
$j\to i$ corresponds to an improvement of the prediction, due to the
knowledge of $\theta_j(t-1)$,$\ldots$,$\theta_j(t-m)$. According to
the theory of Fourier series, predictions can be made performing
linear regression in the feature space spanned by the variables
\begin{equation}\label{c}
{\cal C}_{\vec{\ell},\vec{\delta}}(t)=cos[\ell_1 \theta_1
(t-\delta_1)+\cdots+\ell_n \theta_n (t-\delta_n)], \end{equation}
and
\begin{equation}\label{s}
{\cal S}_{\vec{\ell},\vec{\delta}}(t)=sin[\ell_1 \theta_1
(t-\delta_1)+\cdots+\ell_n \theta_n (t-\delta_n)], \end{equation}
for all integers $\vec{\ell}=(\ell_1,\ldots,\ell_n)\in \mathbb{Z}^n$
and delays $\vec{\delta} \in \{1,\ldots,m\}^n$ (only terms with
$|\ell_1|+\cdots+|\ell_n|<L$ are actually taken into account, $L$
depending on the amount of data at disposal). All the
$N$-dimensional vectors $\{{\cal C},{\cal S}\}$  and
$\{\Delta\theta\}$ can be assumed to have zero mean and unit norm,
after suitable linear transformations. Now we consider the
evaluation of the causality $C(j\to i)$.  We denote $H\subseteq
\Re^N$  the linear span of all vectors $\{{\cal
C}_{\vec{\ell},\vec{\delta}},{\cal S}_{\vec{\ell},\vec{\delta}}\}$
and $H_0\subseteq \Re^N$  the linear span of those vectors $\{{\cal
C}_{\vec{\ell},\vec{\delta}},{\cal S}_{\vec{\ell},\vec{\delta}}\}$
with $\ell_j = 0$. Decomposing $H=H_0\oplus H^\perp$, we denote
$\{u_\alpha\}$ the orthonormal basis of $H^\perp$ \cite{orto} and
calculate
\begin{equation}\label{caus}
C(j\to i)=\sum_{\alpha^\prime}  \left(u_{\alpha^\prime}\cdot
\Delta\theta_i\right)^2 , \end{equation} the sum above being over
significative projections \cite{signi}. $C(j\to i)$ probes the flow
of information from $\theta_j$ to $\theta_i$.

In the following we will show the application of the proposed
analysis to data arising from the Euler discretization of noisy
Kuramoto's equations:
\begin{equation}\label{kuramoto}
\dot{\theta_i}=\omega_i + \beta \sum_{j=1}^n s_{ji}\;
sin\left(\theta_j-\theta_i \right) + \xi_i(t),
\end{equation}
where $s_{ji}$ is equal to one (zero) if there is (not) a coupling
from $\theta_j$ to $\theta_i$. Therefore, we take $L=2$ and $m=1$.

As it has been pointed out in \cite{ding}, a fundamental property of
causality estimators is the ability to discern whether the influence
between two channels is direct or mediated. Therefore, as the first
example, we consider a system of three coupled oscillators in which
the first oscillator is coupled to the second and the second to the
third (no coupling between the first and the third oscillators).
Although there is a relevant phase correlation between the first and
the third variables,
\begin{equation}\label{phcorr}
R_{13}=\left|{1\over N} \sum_{t=1}^N
e^{i(\theta_1(t)-\theta_3(t))}\right|,
\end{equation}
our approach correctly reveals that the influence between the
two time series is actually mediated by the second oscillator (see
figure 1).

As a second example, we consider the transient phase dynamics of a
fully connected system of three oscillators. We fix the values of
couplings so that, in the absence of noise, full synchronization of
the three phases would arise in the large time limit.  At $t=1$ the
phases are randomly assigned by uniform sampling in $[0,2\pi]$.
Subsequently the system undergoes a transient process during which
the phases organize and the cross-trial phase correlation between
oscillators, $R_{ij}(t)=|\langle
e^{i(\theta_i(t)-\theta_j(t))}\rangle|$ where the average is over
initial phases, increases (figure 2-top); the asymptotic value of
$R$ is not exactly one due to the presence of noise. In figure
2-bottom the same process is described in terms of the causality
between oscillators (note that in this case the different samples,
needed to evaluate causality at fixed time, arise from many
realizations of the process with varying initial phases). We observe
that the causality is larger at beginning, when oscillators are
organizing. The value of the causality at large times is much
smaller. In the absence of noise causalities would vanish, at large
$t$: here noise frequently perturbs the system and drives it out of
the synchronized state. A similar cross-trial analysis, in terms of
causalities, may also be performed to study the transient after a
stimulus \cite{tass}.

We consider now the case of a binary matrix $s_{ji}$ corresponding
to an undirected graph, made of $50$ nodes and $50$ links, built by
means of the preferential attachment procedure \cite{pa}: typically,
in these networks there are few nodes with a large number of
connections (hubs), whilst most of the nodes have small
connectivity. We assume that all oscillators in the network have the
same natural frequency $\omega_i =\omega$: we find that the average
phase synchronization between one node and one of its neighbors is
almost independent of the connectivity $k$ (figure 3-top). In figure
3-middle we depict the sum of the outgoing causalities from a node,
as a function of the number of its links $k$: it tends to saturate.
In figure 3-bottom we depict the sum of the incoming causalities of
a node as a function of the connectivity $k$: using the
interpretation of causality in terms of information, these plots
show that nodes with $k>4$ receive more information than they
transmit and suggest the presence of a maximum amount of information
that a node can transmit in this system. This result can be seen
from the point of view of the {\it law of diminishing marginal
returns} \cite{law}, which states that when the amount of a variable
resource is increased, while other resources are kept fixed, the
resulting change in the output will eventually diminish. In the
language of social networks, this is due to the fact that there is a
limit to the information a person may handle \cite{dunbar}. Turning
to consider an undirected graph with $50$ nodes, each connected to
$4$ randomly chosen nodes, we assign different natural frequency at
each node: as displayed in figure 4, we find that both the average
phase synchronization with neighbors and the average causality with
neighbors are nearly independent of $\omega_i$.

As a further application, we consider now a model of interacting
thalamocortical neuronal ensembles proposed in \cite{sheeba} to
account for the behaviour of $\delta$ and $\theta$ waves during
anaesthesia. The model consists of three ensembles of Kuramoto's
oscillators, namely the cortical (CO), the thalamocortical relay
neurons (TC) and the thalamic reticular neurons (RE), having
different mean natural frequencies and characterized by
intra-ensemble and inter-ensemble coupling parameters; CO neurons
receive sensory inputs from TC neurons, TC neurons receive sensory
inputs from both CO and RE neurons, whilst RE neurons receive
sensory inputs only from TC neurons. In order to simulate the
transition from the deeply to the lightly anaesthetized state, the
intra and inter couplings are swept linearly to mimic the effect of
decreasing concentration of anaesthetic agent. In \cite{sheeba} this
model has been studied by analyzing the mean frequencies of
ensembles and the phase correlations between ensembles. In figure 5
we depict the phase correlations and the causalities among the three
groups of neurons, as a function of $\beta$, the overall factor of
couplings. Figure 5-top is in agreement with the results reported in
\cite{sheeba}, whilst figures 5-middle and 5-bottom show that the
transition from deep to light anaesthesia is characterized by
asymmetries in the causality relationships: CO neurons drive TC
neurons, and TC neurons drive RE neurons. These findings complete
the analysis reported in \cite{sheeba}, whose aim was to tackle the
problem of anaesthetic awareness.

Summarizing, we have proposed use of Granger causality for the
analysis of systems of circular variables. Processing data from
simulated Kuramoto systems, we have shown that causality (i)
discerns direct and mediated interactions and (ii) is suitable to
study transient phenomena. Our results, on systems living on
networks built by preferential attachment, on one side support the
interpretation of Granger causality in terms of flow of information,
on the other hand they show that in these systems there is a maximum
amount of information that an oscillator can handle, in accordance
with the law of diminishing marginal returns. Finally, we have
analyzed  a recently proposed model of the transition from deep to
light anaesthesia, and used causality to put in evidence the
drive-response relationships between ensembles. We believe that the
proposed approach will be useful to analyze real data in the form of
phases.

\begin{figure}[ht!]
\begin{center}
\epsfig{file=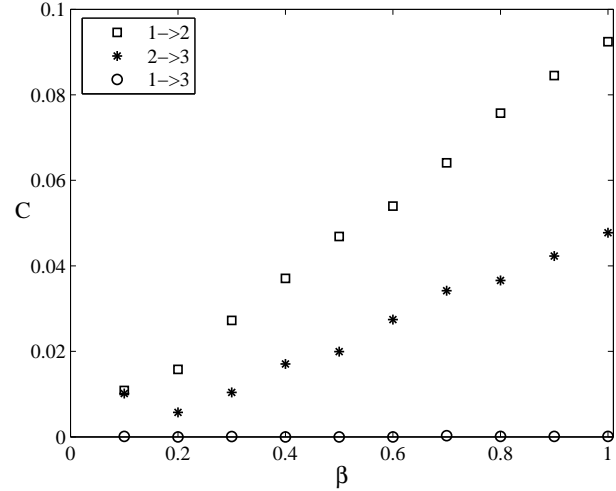,height=7.cm}
\end{center}
\caption{{\small Causalities in a system of three coupled
oscillators, with natural frequencies $\omega_1 =0.3$, $\omega_2
=0.4$, $\omega_3 =0.5$. The noise term has autocorrelation $\langle
\xi_i(t)\xi_j(t^\prime)\rangle=2D\delta_{ij}\delta(t-t^\prime)$,
with $D=0.01$. Couplings with strength $\beta$ are introduced from
oscillator $1$ to oscillator $2$ and from $2$ to $3$. Our estimate
of the causalities are performed on time series of length $N=200$,
recorded in the steady state after the transient. The interaction $1
\to 3$ is correctly recognized as mediated by oscillator $2$. These
results are robust to changes in parameters of the
model.\label{fig1}}}
\end{figure}

\begin{figure}[ht!]
\begin{center}
\epsfig{file=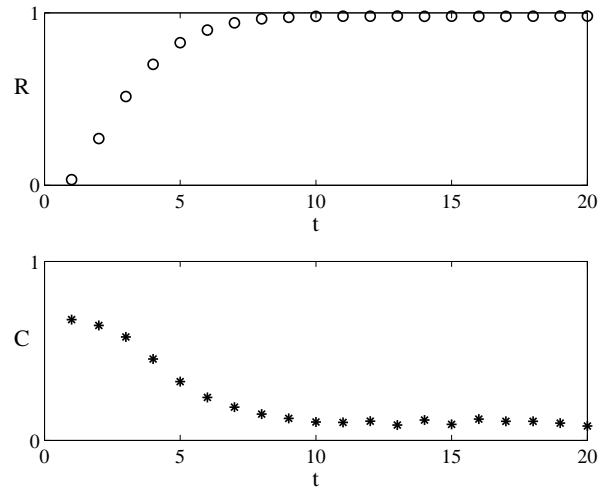,height=7.cm}
\end{center}
\caption{{\small Phase correlation and causality, as a function of
time, for a system of three oscillators, with natural frequencies
$\omega_1 =0.25$, $\omega_2 =0.2$, $\omega_3 =0.15$, noise strength
$D=0.01$ and all-to-all couplings of strength $\beta=0.3$. Different
samples, at fixed $t$, arise from 500 different runs of the model
with random initial phases. The values of $R$ and $C$, here plotted,
are averaged over all pairs of oscillators.  \label{fig2}}}
\end{figure}

\begin{figure}[ht!]
\begin{center}
\epsfig{file=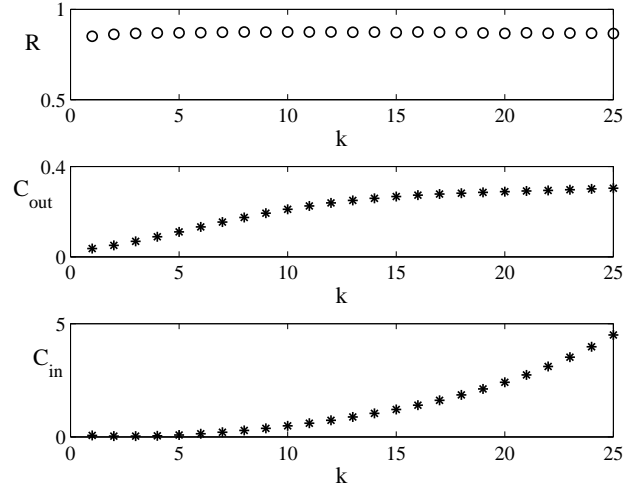,height=7.cm}
\end{center}
\caption{{\small Kuramoto systems living on undirected graphs built
by preferential attachment, are analyzed in terms of causality.
Graphs have 50 nodes and 50 links, all oscillators have the same
natural frequency $\omega =0.4$, couplings $\beta =0.3$ are set
between oscillators connected by a link. The noise term is $D=0.1$.
Quantities refer to the steady state, after the transient. Results
are averaged over 20000 different networks.
 (Top) The mean phase correlation between a node and its neighbors is displayed
as a function of $k$, the number of links of that node. It appears
to be independent of $k$ (Middle) For a node of connectivity $k$,
the sum of outgoing causalities from that node to its neighbors,
$C_{out}$, is depicted. It saturates at large $k$. (Bottom) For a
node of connectivity $k$, the sum of incoming causalities from its
neighbors to that node, $C_{in}$, is depicted. For $k <4$
($k>4$),$C_{in} <C_{out}$ ($C_{in} >C_{out})$. These results are
robust to changes in parameters of the model.\label{fig3}}}
\end{figure}

\begin{figure}[ht!]
\begin{center}
\epsfig{file=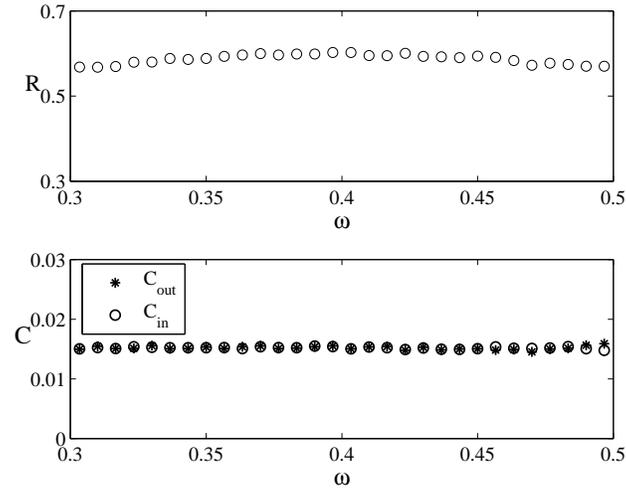,height=7.cm}
\end{center}
\caption{{\small Phase correlation (top) and causality (bottom) are
plotted as a function of the natural frequency $\omega$ for a
Kuramoto system living a random graph of 50 nodes, each node being
connected to 4 other random nodes. Coupling is $\beta =0.05$ and
$D=0.1$. Quantities refer to the steady state, after the transient.
Results are averaged over 10000 different networks. Both $R$ and $C$
appear to be almost independent of $\omega$. \label{fig4}}}
\end{figure}

\begin{figure}[ht!]
\begin{center}
\epsfig{file=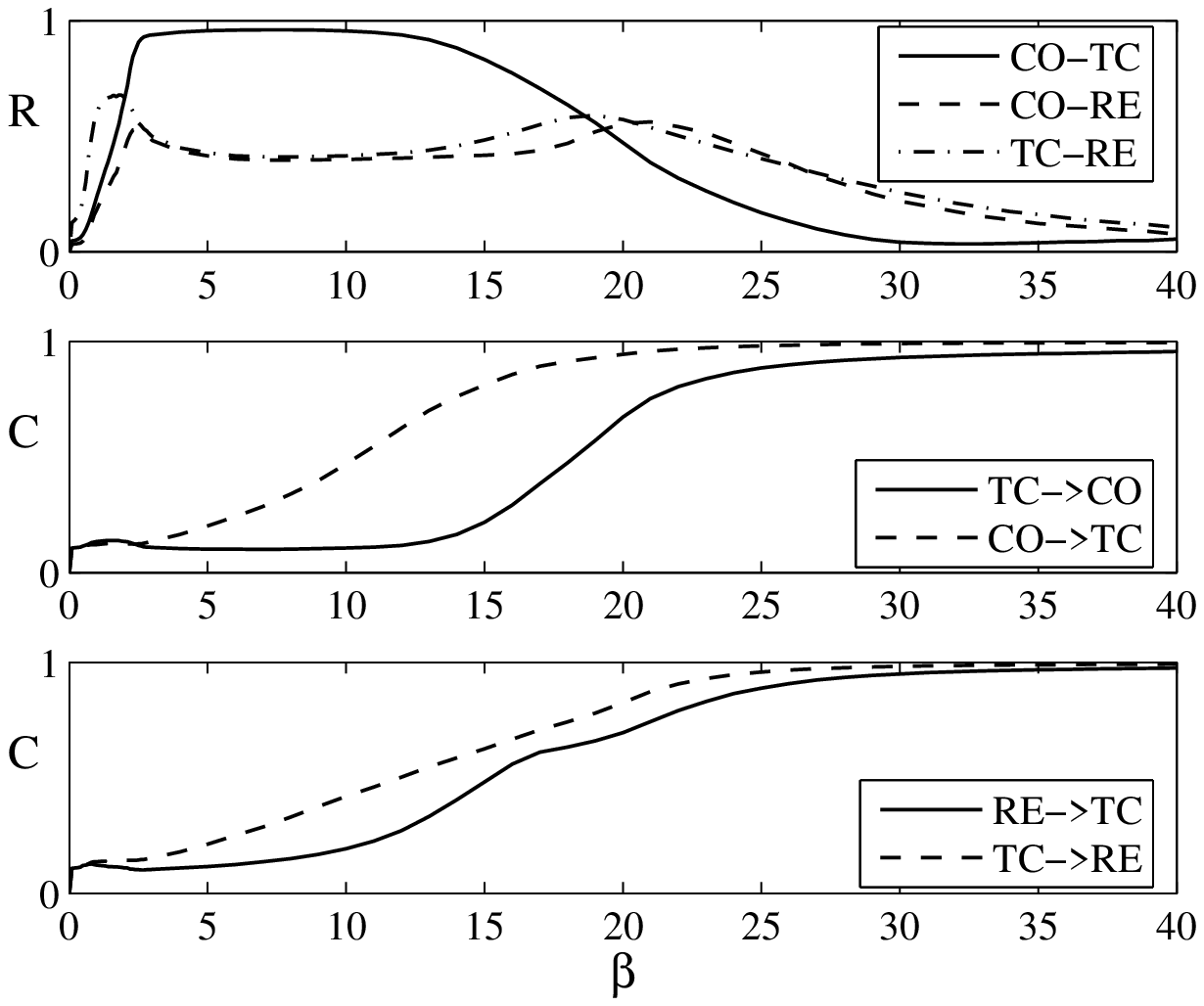,height=7.cm}
\end{center}
\caption{{\small A model to describe the transition from deep to
light anaesthesia is simulated. It consists of three groups of
neurons, CO-TC-RE, each made of 6 oscillators. The parameters
(natural frequencies, inter and intra couplings) are fixed as in
\cite{sheeba}. As a function of an overall factor $\beta$
(multiplying all couplings and thus simulating the effect of
decreasing concentration of anaesthetic agent), the phase
correlation between groups is depicted (top), the causalities
between CO and TC groups (middle) and causalities between TC and RE
groups (bottom). The causality from group A to group B is estimated
as follows. For each oscillator in B, we estimate the improvement in
prediction due to the inclusion, in the regression model, of all the
oscillators in A; then we average the outcome over all the
oscillators in B. \label{fig4}}}
\end{figure}


\begin{thebibliography}{99}
\bibitem{winfree}A.T. Winfree, {\em The geometry of biological time}. (Springer, New York, 1980);Y. Kuramoto, {\em Chemical oscillations, Waves and Turbulence}. (Springer, Berlin, 1984);A. Pikovsky, M. Rosenblum, J. Kurths, {\em Synchronization: A Universal Concept in Nonlinear Sciences}.
(Cambridge University Press, Cambridge, England, 2001).
\bibitem{bocca}S. Boccaletti, J. Kurths, G. Osipov, D.L. Valladares and C. Zhou, Physics Reports {\bf 366}, 1 (2002).
\bibitem{rosen} M.G. Rosenblum, A. Pikovsky, Phys. Rev. E {\bf 64}, 45202R
(2001); D.A. Smirnov, M.B. Bodrov, J.L. Perez Velazquez, R. A.
Wennberg, B. P. Bezruchko, CHAOS {\bf 15}, 024102 (2005).
\bibitem{oscillatornetworks}S. Boccaletti, V.Latora, Y.Moreno, M.Chavez and D.U.Hwang, Physics Reports {\bf 424}, 175-308 (2006);
M. Brede, Phys. Lett. A {\bf 372}, 2618 (2008).
\bibitem{hla} C.W.J. Granger, Econometrica {\bf 37}, 424 (1969);
for a review see K. Hlavackova-Schindler, M. Palus, M. Vejmelka, J.
Bhattacharya, Physics Reports {\bf 441}, 1 (2007).
\bibitem{chen} Y. Chen, G. Rangarajan, J. Feng and M. Ding, Phys. Lett. A {\bf 324}, 26 (2004).
\bibitem{blinoska} K.J. Blinowska, R. Kus, M. Kaminski, Phys. Rev. E {\bf 70},
50902(R) (2004).
\bibitem{seth} A.K. Seth, Network: Computation in
Neural Systems {\bf 16}, 35 (2005).
\bibitem{faes}L. Faes, A. Porta, G. Nollo, Phys. Rev. E {\bf 78}, 026201
(2008).
\bibitem{schreiber} Transfer entropy, introduced in T. Schreiber, Phys. Rev. Lett. {\bf 85}, 461 (2000), measures the flow of Shannon information between two time series.
In Marinazzo et al., Phys. Rev. Lett. {\bf 100}, 144103 (2008), it
was shown that Granger causality implies non-zero transfer entropy.
\bibitem{noi}D. Marinazzo, M.
Pellicoro and S. Stramaglia, Phys. Rev. E {\bf 77}, 056215 (2008).
\bibitem{orto} The basis $\{u\}$ consists of the eigenvectors, with
non-vanishing eigenvalue, of the matrix $K=X X^\top -P_0 X X^\top -X
X^\top P_0 + P_0 X X^\top P_0$, where $X$ is the matrix having all
vectors $\{{\cal C}_{\vec{\ell},\vec{\delta}},{\cal
S}_{\vec{\ell},\vec{\delta}}\}$ as columns and $P_0$ is the
projection matrix onto $H_0$.
\bibitem{signi} Firstly we observe that $u_\alpha \cdot \Delta\theta_i=u_\alpha \cdot \Omega$, where
$\Omega=\Delta\theta_i-P_0\Delta\theta_i$. Let us call
$\Omega^\prime= \Omega/\sqrt{\Omega\cdot\Omega}$. For large $N$, and
under the null hypothesis of statistical independence between
$u_\alpha$ and $\Omega^\prime$, the quantity $u_\alpha \cdot
\Omega^\prime$ can be treated as a Gaussian variable with variance
1/N. This allows the selection of significative projections, with
$99.5 \%$ confidence, by means of the Bonferroni approach for
multiple tests.
\bibitem{ding}M. Kaminski, M. Ding, W. Truccolo, and S.L. Bressler, Biological Cybernetics
{\bf 85}, 147 (2001).
\bibitem{tass} P.A. Tass, Phys. Rev. E {\bf 67}, 051902
(2003).
\bibitem{pa} A.L. Barabasi, R. Albert, Science {\bf 286}, 509
(1999).
\bibitem{law} L. Lopez and M.A.F. Sanjuan, Phys. Rev. E {\bf 65}, 036107 (2002).
\bibitem{dunbar}R.I.M. Dunbar, Behav. Brain Sci. {\bf 16}, 681 (1993).
\bibitem{sheeba} J.H. Sheeba, A. Stefanovska, P.V.E. McClintock,
Biophys. J. {\bf 95}, 2722 (2008).
\end{thebibliography}
\end{document}